\documentclass[12pt]{article}
\usepackage{graphicx}
\usepackage{float}
\usepackage{graphicx,caption}
\usepackage{xcolor}
\usepackage{verbatim}
\usepackage[margin=0.75in]{geometry}
\usepackage{amsmath,amssymb,amsfonts, braket}
\usepackage{authblk}

\newcommand{\er}{Er$^{3+}$}
\newcommand{\ercawo}{Er$^{3+}$:CaWO$_4$}
\newcommand{\cawo}{CaWO$_4$}
\newcommand{\w}{$^{183}$W}

\title{Coherent control of interacting solid-state spins below the diffraction limit}

\begin{document}

\author[1]{Haitong Xu\thanks{These authors contributed equally to this work.}}
\author[1]{Mehmet T. Uysal\textsuperscript{*}}
\author[1]{\L{}ukasz Dusanowski\thanks{Present address: Department of Electrical and Computer Engineering, FAMU-FSU College of Engineering, Florida State University, Tallahassee, FL 32310, USA}}
\author[1]{Adam Turflinger}
\author[1]{Ashwin K. Boddeti}
\author[1]{Joseph Alexander}
\author[1]{Jeff D. Thompson\thanks{jdthompson@princeton.edu}}
\affil[1]{Department of Electrical \& Computer Engineering, Princeton University, Princeton, NJ 08544, USA}

\maketitle

\textbf{Optically addressed atomic defects in the solid-state are widely used as single-photon sources and memories for quantum network applications. 
The solid-state environment allows for a high density of electron and nuclear spins with the potential to form registers for coherent information processing. 
However, it is challenging to reliably address individual spins at nanometer separations where interactions are large. 
Rare-earth ions offer a unique solution, as their narrow homogeneous optical linewidth allows frequency-domain resolution of a large number of emitters without regard to their spatial separation. 
In this work, we realize coherent optical and spin control of a pair of interacting \er{} ions, together with a nearby nuclear spin ancilla. 
We demonstrate two-qubit electron-electron gates and use them to perform repeated quantum non-demolition measurements on one of the \er{} ions. 
We also demonstrate electron-nuclear gates to allow coherent storage and retrieval of qubit information in a nuclear spin, and show that the nuclear spin coherence survives readout of the electron spin. 
These techniques can be readily scaled to larger numbers of electron and nuclear spins, paving the way for massively multiplexed quantum network nodes. }

\section{Introduction}
Spin-photon entanglement using optically addressable solid-state atomic defects is a key ingredient for scalable quantum networks~\cite{Togan2010,Bernien2013,Knaut2024,inc2024distributedquantumcomputingsilicon,Fang2024,Ruskuc2025,uysal2025}.
Interactions between a central electron and nearby nuclear spins can be used to implement long-lived ancilla memories \cite{Dutt2007,neumann2010single,Maurer2012,Bourassa2020,Higginbottom2023, song2025}, enabling sophisticated networking protocols~\cite{Kalb2017, Pompili2021, Hermans2022}.
Multi-qubit nuclear spin registers can be controlled using gradients in the hyperfine coupling \cite{Zhao2012,Taminiau2012,Kolkowitz2012,Bradley2019, Abobeih2022}, and similar techniques can be extended to control ``dark" electron spins surrounding a single NV center \cite{Gaebel2006, Degen2021}.
In all of these approaches, control and measurement are performed using a single, central electron spin, which limits the achievable register size.
This limit has been partially overcome using sub-diffraction-limit addressing of interacting NV centers based on their crystallographic orientation \cite{Neumann2010, Dolde2013, Lee2023}, but this technique cannot scale to many spins.

Frequency-domain addressing is an alternative approach to sub-diffraction-limit control, relying on natural or engineered variations in the resonance frequency of individual emitters.
This technique is particularly applicable for rare-earth ion defects, which have very narrow homogeneous optical linewidths.
This capability has been used for multimode classical \cite{Mossberg1982} and quantum \cite{Afzelius2010dem,Zhong2017,Yang2018,seri2019} memories in large ion ensembles, and for resolving and manipulating individual ions below the diffraction limit \cite{zhong2018,dibos2018,Chen2020,Ulanowski2022}.
A number of theoretical proposals have considered the use of individually resolvable interacting ions for coherent quantum information processing \cite{wesenberg2007,kinos2021}.
Despite recent advances in coherent control of individual rare-earth ions \cite{Chen2020,Kindem2020,ourari2023indistinguishable,Wang2023} and their interactions with nearby nuclear spins \cite{Ruskuc2022,uysal2023coherent,Travesdo2025}, interactions between individually controlled rare-earth ion electron spins have not been demonstrated. 

In this work, we demonstrate individual coherent control over two interacting \er{} spins in \cawo{}, separated by tens of nanometers inside a single photonic crystal cavity.
Targeted optical and spin control is enabled by inhomogeneous broadening of the optical transition frequencies and spin g-factors, respectively.
We leverage magnetic dipolar interactions between the ions to realize electron-electron gates and create a Bell pair with fidelity $F=0.76(2)$.
We also implement gates between one of the electrons and a nearby \w{} nuclear ancilla, using a novel optimal control technique, which we use to characterize the nuclear spin coherence time ($T_{2} = 1.0(2)$s) and demonstrate coherent storage and retrieval of information in the nuclear spin.
Finally, we show that the nuclear ancilla coherence is preserved while measuring the spin of the remote electron, demonstrating a pathway to multi-mode quantum repeaters in this platform.
We discuss future extensions to more than two interacting electron spins, and the implementation of multiplexed quantum repeater nodes.

\section{Results}
\begin{figure}[H]
    \makebox[\textwidth][c]{\includegraphics[]{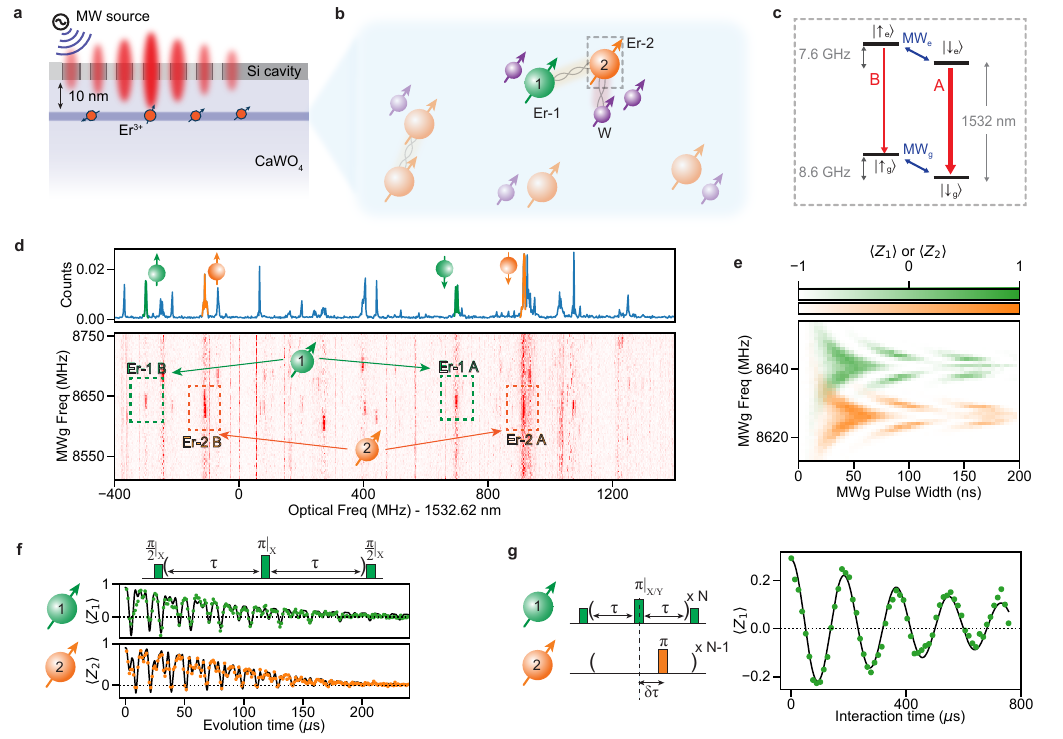}}%
    \caption{
    Spectroscopy of an interacting \er{} ion pair
    \textbf{a,} Our devices consist of a \cawo{} substrate with a shallow implanted \er{} layer, coupled to a silicon nanophotonic cavity providing an optical interface. A microwave (MW) antenna provides coherent spin control.
    \textbf{b,} We study an interacting pair of \er{} ions and their proximal nuclear spins (\w{}, with spin $I=1/2$).
    \textbf{c,} Level diagram of \ercawo{}, indicating splittings of the ground and excited state ($S=1/2$) manifolds in an external field of 790 Gauss. The field is oriented near the crystal a-axis to optimize Purcell enhancement, $(\theta,\phi)=(100^\circ,95^\circ)$ \cite{Raha2020}.
    \textbf{d,} The PLE spectrum shows separated optical transition frequencies of the \er{} ions in the cavity (top panel). MW-assisted PLE spectroscopy further reveals different ground state spin transition frequencies (bottom panel). The ions studied in this work, Er-1 and Er-2, are indicated in each panel.
    \textbf{e,}  Z basis expectation in MWg manifold for Er-1 ($\langle Z_{1}\rangle$, green) and Er-2 ($\langle Z_{2}\rangle$, orange) after a MW pulse, revealing separated chevron patterns and enabling selective spin control.
    \textbf{f,} Hahn echo spin coherence measurements on both ions. The data is fitted to an electron spin echo envelope modulation (ESEEM) model incorporating three nearby \w{} nuclear spins (Extended Tab.~\ref{Tab:SI_NucProperties}) with a stretched exponential decay envelope $e^{-(2\tau/T_2)^n}$ (solid lines). The fit for Er-1 yields $T_2= 100(6)\,\mu \textrm{s}$ and $n=1.1(1)$, while the fit to Er-2 yields $T_2 = 120(2)\,\mu \textrm{s}$ and $n=2.0(1)$.
    \textbf{g,} To allow the ions to interact while protecting from decoherence, we apply XY-N decoupling sequences on Er-1 and simultaneously apply $\pi$-pulses to flip Er-2. The effective interaction time is given by $2(N-1)(\tau-\delta\tau)$, where $\delta\tau$ is the delay between the $\pi$-pulses on each ion (here, $N=64$). A coherent interaction strength of $J=5.40(3)$~kHz is determined from an exponentially damped oscillation fit to the data.
}
    \label{fig:Fig1}
\end{figure}

Our device architecture is illustrated schematically in Fig.~\ref{fig:Fig1}a,b and previously described in Ref.~\cite{Chen2021a,ourari2023indistinguishable}.
In brief, \er{} ions implanted $\sim$10 nm below the surface of a \cawo{} crystal couple evanescently to a silicon nanophotonic device.
Experiments are performed in a $^3$He cryostat at a temperature of $500$~mK, with fiber-coupled input and output.
Under an external magnetic field $B_{\text{ext}}=790$~Gauss, the \er{} ground and excited state spin manifolds are split by 8.6~GHz and 7.6~GHz respectively to form the 4-level system (Fig.~\ref{fig:Fig1}c).
The spin-conserving optical transitions, $A$ and $B$, are selectively enhanced by the optical cavity (with Purcell factor $P \approx 200$ for the ions in this work), enabling single-shot spin readout with fidelity $0.94$ \cite{Kindem2020,Raha2020}.

The optical transitions of many ions in the single-mode cavity can be resolved within the inhomogeneous linewidth (Fig.~\ref{fig:Fig1}d, top).
A unique property of \ercawo{} is that there is also significant disorder in the magnetic moment (attributed to the $S_4$ site symmetry~\cite{ourari2023indistinguishable}), which gives rise to inhomogeneous broadening of the spin transitions and further enables individual spin control using microwaves. 
Using microwave-assisted photoluminescence excitation spectroscopy (PLE) -- exciting the ions with a resonant laser and remixing spin population from dark states with microwaves -- we can simultaneously map the optical and spin transition frequencies for all ions within a cavity (Fig.~\ref{fig:Fig1}d, bottom).
The ions selected for this work (Er-1 and Er-2, Fig.~\ref{fig:Fig1}b), have optical transitions separated by 200 MHz, and spin transitions separated by 14 MHz (0.16\% difference in magnetic moment), which is sufficient to selectively control both optical and spin transitions on each ion (Fig.~\ref{fig:Fig1}e).
The Hahn echo coherence time of the ions is 100(6) and 120(2) $\mu$s, and the echo envelope exhibits strong modulation from local \w{} nuclear spins (Fig.~\ref{fig:Fig1}f). 
Using double electron-electron resonance spectroscopy (DEER) with an inter-pulse spacing chosen to avoid resonant coupling with the \w{} bath, we measure a magnetic dipolar interaction strength of $J=5.40(3)$ kHz between the two electron spins (Fig.~\ref{fig:Fig1}g).
This strength implies a separation between the spins of 10 -- 42 nm, depending on their orientations (SI.~\ref{SI:Er_search}).

Next, we leverage the interaction between the ions to implement an entangling gate. 
An electron-electron controlled-Z (CZ) gate is constructed using interleaved dynamical decoupling sequences on the two ions to accumulate a two-qubit phase while also decoupling from slowly-varying local magnetic fields and the \w{} nuclear spin bath (Fig.~\ref{fig:Fig2}a).
By offsetting the decoupling sequences on the two ions, the effective interaction time $T_{\text{int}}$ can be controlled independently of the sequence period (SI.~\ref{SI:er_gate}).
We characterize the gate performance by the resultant Bell pair fidelity, obtained by measuring the correlations in three orthogonal bases,  $F=(1+\langle X_1X_2\rangle+\langle Y_1Y_2\rangle+\langle Z_1Z_2\rangle)/4$ (Fig.~\ref{fig:Fig2}b).
We obtain a raw fidelity of $F=0.66(2)$, which becomes $0.76(2)$ after correcting for readout errors (SI.~\ref{SI:erReadout}).
The CZ gate error arises primarily from single-qubit decoherence during the gate (SI.~\ref{SI:gate_fidelity}).

One application of this gate is remote spin readout: performing a QND measurement of Er-2 by repeatedly transferring its state to Er-1 and measuring Er-1(Fig.~\ref{fig:Fig2}c)~\cite{hume2007,jiang2009}.
A single round of this protocol yields a remote-readout fidelity of 0.88.
Repeating the sequence improves the fidelity to a maximum of 0.92 after three rounds, beyond which accumulated gate errors cause the Er-2 Z-basis population to decay (Fig.\ref{fig:Fig2}d,e).
Although this fidelity is lower than the direct readout fidelity of 0.94, the QND readout protocol allows postselecting runs in which all outcomes agree to achieve 0.98 fidelity after three rounds at the cost of a 66\% acceptance rate~\cite{Kindem2020}.

Next we turn to gate operations between Er-2 and a nearby \w{} nuclear spin.
Under the secular approximation, the electron-nuclear spin evolution is described by the Hamiltonian:
\begin{equation}\label{eq:Nuc_H}
H/\hbar=\ket{\uparrow_g}\bra{\uparrow_g}\otimes\left(\omega_{+}\mathbf{m}_{+}\cdot\mathbf{I}\right)+\ket{\downarrow_g}\bra{\downarrow_g}\otimes\left(\omega_{-}\mathbf{m}_{-}\cdot\mathbf{I}\right),
\end{equation}
where the nuclear spin $\mathbf{I}$ (with $|\mathbf{I}|=1/2$) precesses around the axis $\omega_{\pm}\mathbf{m}_{\pm} = \pm A_{\perp}\hat{\mathbf{x}} +\left(\omega_{L}\pm A_{\parallel}\right)\hat{\mathbf{z}}$ depending on the Er-2 electron spin state. 
Here, $A_\parallel = 2 \pi \times 287 \text{ kHz}, A_\perp = 2 \pi \times 163 \text{ kHz}$ denote the parallel and perpendicular hyperfine coupling strengths, respectively, and $\omega_{L} = 2 \pi \times 142 \text{ kHz}$ is the nuclear spin Larmor frequency in the external magnetic field (Fig.~\ref{fig:Fig3}a, SI.~\ref{SI:nuc_properties}).
This conditional precession allows coherent control of the nuclear spin state by toggling Er-2, which we use to implement two electron-nuclear spin gates, achieving universal two-qubit control by only driving the electron spin.

We first implement an electron-nuclear CZ gate using an XY-2 sequence on Er-2 with inter-pulse spacing resonant with the nuclear spin precession \cite{Zhao2012,Taminiau2012,Kolkowitz2012}. 
The nuclear spin experiences a conditional rotation along an axis $\mathbf{z}'$ by an angle $\pm \alpha$ depending on the initial electron spin state (SI.~\ref{SI:nuc_gate}). 
For our parameters, the XY-2 sequence with an inter-pulse spacing of $\tau=6.208$~$\mu$s results in $\alpha = 0.248\pi$, such that the XY-4 sequence with this spacing nearly ideally realizes a conditional $\pm \pi/2$ rotation around $\mathbf{z}'$ (Fig.~\ref{fig:Fig3}bc). 
We choose $\mathbf{z}'$ as the Z basis of the nuclear spin qubit so that the XY-4 sequence implements a CZ gate.

To achieve full control of the electron-nuclear spin system without the ability to drive direct rotations on the nuclear spin, we need to implement a second controlled rotation gate with an orthogonal axis.
To this end, we introduce a numerical search protocol, GRadient Ascent Sequence Search (GRASS) (SI.~\ref{SI:GRASS}). 
GRASS extends typical evenly-spaced pulse sequences to include arbitrary inter-pulse spacings $\{\tau_1,\ldots,\tau_{N+1}\}$ (Fig.~\ref{fig:Fig3}d).
This parametrization results in a large search space that can be explored using gradient ascent techniques to optimize the overlap between the effective nuclear spin evolution and a desired target rotation.
At the same time, suitable objective functions can also enforce decoupling of the electron from other nuclear spins by only allowing unconditional rotations of the nuclear spin and protect electron spin coherence by decoupling disorder in the environment, such as low frequency magnetic or electric noise \cite{uysal2025}.
After optimizing several target rotations, we choose the shortest sequence that meets our requirements, resulting in a $46.6\,\mu$s sequence consisting of eight‑$\pi$-pulses. This sequence generates a conditional rotation, $CU$, that is equivalent to a controlled-$Y$ gate followed by a Hadamard gate on the nuclear spin (Fig.~\ref{fig:Fig3}d).
 
With these operations, we can measure the coherence of the nuclear spin. 
We find a nuclear spin free precession time of $T_{2}^{*}=34.1(3)$~ms (Fig.~\ref{fig:Fig3}ef), and a Hahn echo coherence time of $T_{2}=1.0(2)$~s (Fig.~\ref{fig:Fig3}gh). 
These numbers improve on the electron spin coherence by four orders of magnitude. 
The improvement on the free precession time is slightly lower than the ratio of their magnetic moments $\mu_e/\mu_n\approx6\times10^4$ because the nuclear spin is strongly interacting with Er-2 (SI.~\ref{SI:nuc_coh}).

The long coherence time of the nuclear spin makes it an ideal ancilla memory for the electron spins. 
To illustrate this functionality, we demonstrate a SWAP gate between the electron and nuclear spin (Fig.~\ref{fig:Fig4}a). 
After using two SWAPs to store an initialized electron state into the nuclear spin and retrieving it, we measure a retrieved state fidelity of $F=0.833(3)$, averaged over the X and Z bases and corrected for electron spin readout (Fig.~\ref{fig:Fig4}b).
The SWAP fidelity is primarily limited by electron spin decoherence over 150~$\mu$s long gate (SI.~\ref{SI:gate_fidelity}).
By using the SWAP gate to initialize the nuclear spin, we can construct an electron-nuclear Bell pair (Fig.~\ref{fig:Fig4}c). 
In this circuit, we use an additional SWAP gate between the two electrons to avoid mid-circuit re-initialization of Er-2, and run a similar circuit in reverse to perform the final state measurement. 
The observed Bell state fidelity after readout correction is $F=0.52(2)$ (Fig.~\ref{fig:Fig4}d), which includes two electron-nuclear SWAPs, two electron-electron SWAPs, and the electron-nuclear CU generating the Bell pair.

A useful ancilla memory must survive multiple operations on the electron spin, including mid-circuit measurement and spin-photon entanglement generation. 
The nuclear spin used in this work is relatively strongly coupled to the Er-2 electron spin ($A_{\parallel},A_\perp \approx \omega_L$), and therefore is strongly perturbed when Er-2 is excited, because of the large change in the electronic magnetic moment between the Er-2 ground and excited state (Fig.~\ref{fig:Fig4}ef). 
While this problem is typically circumvented using weakly coupled nuclear spins~\cite{reiserer2016}, we can instead use the more distant Er-1 electron, which can couple to the nuclear spin indirectly through Er-2. 
In Fig.~\ref{fig:Fig4}g, we show that repeated mid-circuit measurements of Er-1 do not affect the population of the nuclear spin at all, and the coherence is only reduced by 18\% after Er-1 single-shot readout (60 optical pulses) due to Er-2 g-factor fluctuation induced by the incident laser light \cite{uysal2025} (SI.~\ref{SI:nuc_coh}). 

\section{Discussion and conclusion}
This work illustrates the potential for complex multi-qubit operations in a system of individually controllable interacting electron spins. 
We demonstrate two-qubit gates between electron spins, and repeated QND measurements of one electron spin using the other. 
Using the GRASS method to construct arbitrary electron-nuclear gates, we achieve coherent control of a nuclear spin with a second-long coherence time. 
The nuclear spin ancilla can survive repeated optical excitation and measurement of the remote electron spin. This capability has not been previously demonstrated for rare-earth ion qubits, but is an essential ingredient for implementing entanglement swapping and distillation protocols.

The demonstrated gate fidelities are primarily limited by the electron spin coherence, which is in turn believed to be limited by other impurities in the sample~\cite{ourari2023indistinguishable}. By removing or polarizing the impurity bath, the coherence time can be extended by two orders of magnitude~\cite{LeDantec2021}, which should lead to dramatically higher fidelities for both electron-electron and electron-nuclear gates.

Crucially, the addressing and control techniques demonstrated in this work can be extended to larger clusters of interacting spins. 
The optical inhomogeneous linewidth of \ercawo{} (730~MHz) in comparison to the narrow homogeneous linewidths (150 kHz \cite{ourari2023indistinguishable}) allows for resolving hundreds of ions.
The spin inhomogeneous broadening also allows addressing at least ten ions (Fig.~\ref{fig:Fig1}d bottom), and spin control of larger clusters can be achieved by optical AC stark shifts \cite{Chen2020}.
While the density in our current sample is too low to observe large numbers of interacting ions, spatially patterned ion implantation using masks or nano-implantation~\cite{berning2019} may be used to create dense clusters of tens of interacting ions. By further exploiting several nuclear spins around each electron using radio-frequency control~\cite{Bradley2019,Beukers2025}, a total register size of approximately 100 qubits is within reach. Together with the recent demonstrations of direct spin-photon entanglement at the telecom band with \ercawo{} \cite{uysal2025} and remote entanglement with a multi-emitter rare-earth system \cite{Ruskuc2025}, this opens a promising path to highly multiplexed quantum repeater nodes \cite{collins2007multiplexed}, as well as entanglement distillation \cite{Briegel1998} and deterministic photonic cluster state generation \cite{varnava2006loss, borregaard2020} for quantum networks.

\section{Acknowledgments}
We acknowledge helpful conversations with Patrice Bertet, Nathalie de Leon and Jared Rovny. 
This work was supported by the U.S. Department of Energy, under contract number DE-SC0020120 (supporting research on interacting spins), and by the Office of Science, National Quantum Information Science Research Centers, Co-design Center for Quantum Advantage (C2QA) under contract number DE-SC0012704 (supporting device fabrication, materials spectroscopy and improvements).

\newpage

\newpage

\begin{figure}[H]
    \makebox[\textwidth][c]{\includegraphics[]{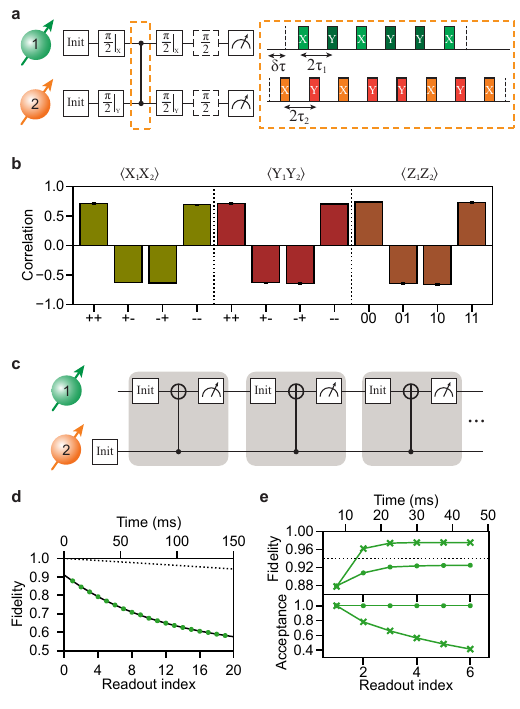}}%
    \caption{
    \er{}-\er{} gates and remote readout.
    \textbf{a,} Bell state preparation and measurement circuit. The $\pi/2$-gates with dashed boxes indicate the analysis pulses for readout in different bases. The CZ gate consists of two sequences with half periods $\tau_{1}=6.220\,\mu$s and $\tau_{2}=6.208\,\mu$s, which are offset by $\delta\tau=2.3\,\mu$s.
    \textbf{b,} Measured correlations in three bases (after correction for readout error), yielding a Bell state fidelity $F=0.76(2)$.
    \textbf{c,} Experimental sequence to read out the spin of Er-2 remotely using Er-1.
    \textbf{d,} Single-shot readout fidelity measured in each remote readout round. The first round readout fidelity of $0.88$ is limited by both two-qubit gate error and direct readout fidelity. Fitting the fidelity of all rounds shows an $8\%$ spin flip probability per round (solid line). In the ideal case with perfect two-qubit gates, the fidelity would be limited by the Er-2 spin relaxation ($T_{1}=2.5(2)$ s, dashed line).
    \textbf{e,} Support Vector Machine (SVM) classifier (circles) or post-selection (crosses) on photon counts show improved fidelity when more rounds are performed and analyzed together. In particular, post-selected readout can achieve higher fidelity than direct optical readout 0.94 on Er-2 (dashed lines), with a 66\% acceptance probability.
}
    \label{fig:Fig2}
    
\end{figure}

\newpage

\begin{figure}[H]
    \makebox[\textwidth][c]{\includegraphics[]{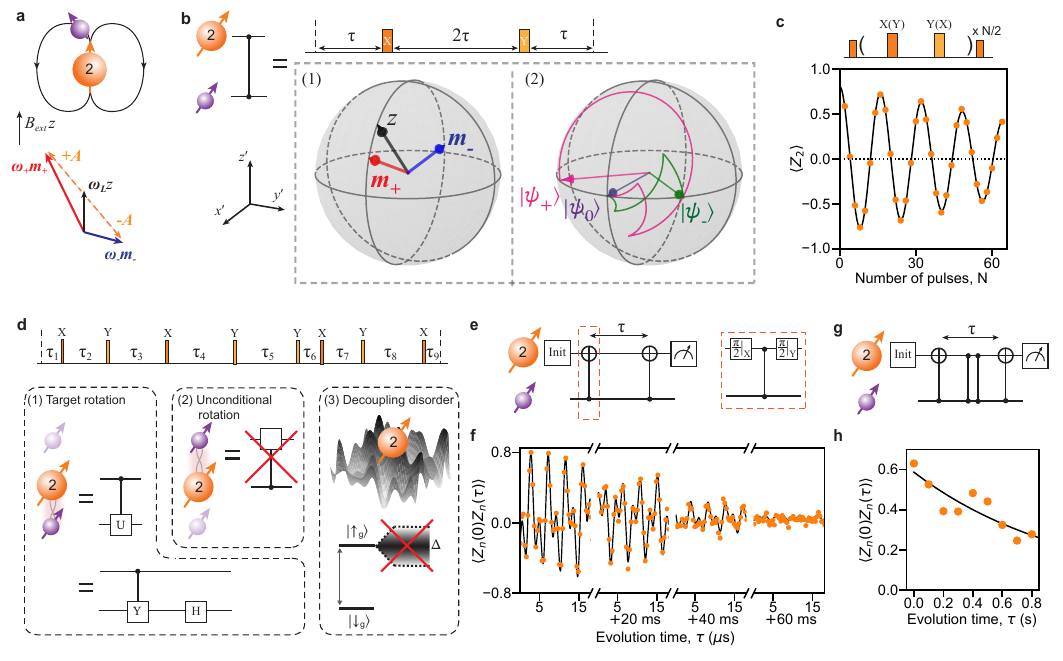}}%
    \caption{
    \w{} nuclear spin gates.
    \textbf{a,} The nuclear spin (purple) precesses around the external magnetic field, $B_{\text{ext}}\mathbf{z}$, and the field from Er-2, $\mathbf{A}=A_\perp\mathbf{x}+A_\parallel\mathbf{z}$, which reverses depending on the direction of the electron spin. This gives rise to electron-state dependent precession axes, $\omega_\pm\mathbf{m}_{\pm}=\omega_L\mathbf{z}\pm\mathbf{A}$.
    \textbf{b,} A CZ gate is constructed from an XY-2 sequence on Er-2 with inter-pulse spacing $\tau=6.208\,\mu$s. The Bloch spheres are orientated along the qubit definition $\mathbf{z}'$ of the nuclear spin. (1) Unit vectors describing the precession axes $\mathbf{m}_{\pm}$, (2) nuclear spin evolution from $\ket{\psi_0}=\ket{+}$ to $\ket{\psi_\pm}=R_{z^\prime}(\pm\alpha)\ket{+}$ depending on initial Er-2 spin state. In each interval, the state precesses around $\mathbf{m}_{\pm}$ for more than one revolution, but only the net rotation is shown for clarity.
    \textbf{c,} By adding additional pulses with the same inter-pulse spacing in \textbf{b}, a larger phase accumulation can be observed. The oscillation is fitted to extract the nuclear spin rotation angle $\alpha=0.248\pi$ every two pulses (solid line).
    \textbf{d,} Pulse sequences with arbitrary inter-pulse spacing can be used to implement a desired two-qubit operation by optimizing for three goals: (1) achieving a target rotation on the objective nuclear spin, (2) preventing unwanted coupling to other nuclear spins by only allowing unconditional rotations, and (3) improving electron spin coherence by decoupling from disorder. The optimization results in the target controlled-unitary CU, which is equivalent to a CY gate with a Hadamard gate on the nuclear spin.
    \textbf{e,} To measure nuclear spin coherence, two CX gates (controlled by nuclear spin) constructed from CZ gates (from \textbf{b}) are performed to measure $Z$ correlation on the nuclear spin $\langle Z_{n}(0)Z_{n}(\tau)\rangle$. The nuclear spin precession axes $\mathbf{m}_{\pm}$ are almost perpendicular to $\mathbf{z}'$ (as in \textbf{b}), and thus this correlation measures nuclear spin free precession.
    \textbf{f,} A correlation experiment (Ramsey) reveals free precessions at frequencies $\omega_{\pm}$ with the fitted coherence time $T_{2}^{*}=34.1(3)$~ms (solid line).
    \textbf{g,} Two CZ gates in the middle of the correlation experiment in \textbf{e} act as a refocusing operation for the Hahn experiment.
    \textbf{h,} Nuclear spin Hahn measurement shows fitted $T_{2}^{\text{Hahn}}=1.0(2)$~s (solid line).
    }
    \label{fig:Fig3}
\end{figure}

\newpage

\begin{figure}[H]
    \makebox[\textwidth][c]{\includegraphics[]{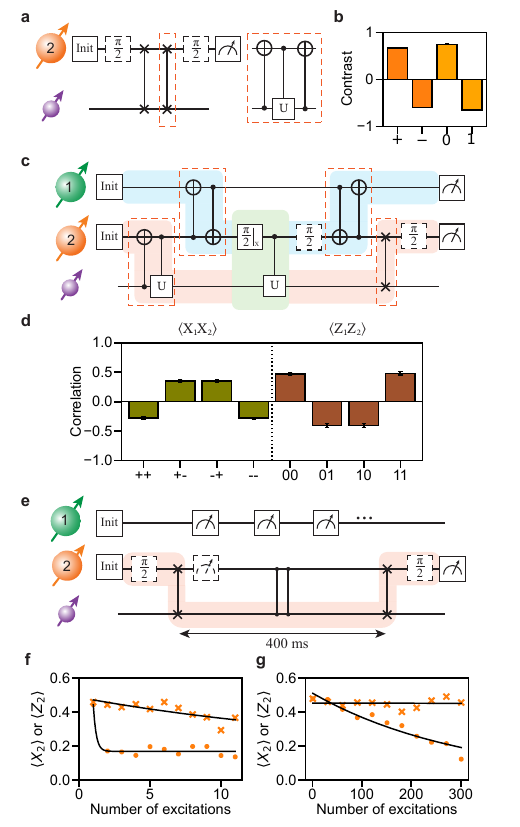}}%
    \caption{
    Nuclear spin quantum memory.
    \textbf{a,} A SWAP gate can be constructed by three CX and CU operations and is used to store and retrieve a qubit.
    \textbf{b,} Measured contrast when storing and retrieving $\ket{0/1}$ or $\ket{\pm}$ states shows $\bra{0}\text{SWAP}^2\ket{0/1}=0.67(1), -0.60(1)$ and $\bra{+}\text{SWAP}^2\ket{\pm}=0.75(1),-0.65(1)$.
    \textbf{c,} Circuit to generate and measure a Bell state between Er-2 and the nuclear spin, consisting of four SWAPs (red dashed boxes) between Er-1 and Er-2 or Er-2 and the nuclear spin. Three of the SWAPs contain only two CX (CU) operations because only Z basis information is needed (SI.~\ref{SI:nuc_circuit}).
    The $\pi/2$ operations in dashed boxes are optional to measure in different bases.
    \textbf{d,} Correlating the readouts on Er-1 and Er-2 after the circuit in \textbf{c} yields averaged expectations, $|\langle X_1X_2\rangle|=0.31(2)$ and $|\langle Z_1Z_2\rangle|=0.44(3)$, revealing Bell state fidelity $F=0.52(2)$ between Er-2 and the nuclear spin.
    \textbf{e,} Information is swapped and stored in the nuclear spin for 400~ms to show its quantum memory capability, while one of the \er{} ions is optically excited to mimic a mid-circuit readout, with each consisting of 60 optical pulses. Then the qubit is retrieved and measured through Er-2.
    \textbf{f,} Optical excitations of Er-2 immediately decoheres the nuclear spin with 1 excitation ($\langle Z_2\rangle$, circles) and induces spin relaxation with a fitted exponential decay constant of 20 pulses ($\langle X_2\rangle$, crosses). 
    Z basis is less robust because the nuclear spin precession axes $\mathbf{m}_{\pm}$ are almost perpendicular to $\mathbf{z}'$ (Fig.~\ref{fig:Fig3}b).
    \textbf{g,} Optical excitation of Er-1 slightly decoheres the nuclear spin ($\langle Z_2\rangle$, circles), with fitted decaying constant of 307 pulses. No spin relaxation is observed ($\langle X_2\rangle$, crosses).
}
    \label{fig:Fig4}
\end{figure}

\newpage

\setcounter{figure}{0}
\renewcommand{\figurename}{Extended Figure}
\renewcommand{\tablename}{Extended Table}

\section{Supplementary information}
\subsection{Characterizing interacting \er{} pairs}\label{SI:Er_search}
Inhomogeneous broadening of both spin and optical transitions allows for individual ion addressing in the optical and microwave domain. 
Inhomogeneous broadening of the optical transition is common in solid-state emitters, and arises from local environment variations, \emph{i.e.} local defects or strain, and the typical optical inhomogeneous linewidth in \ercawo{} is 730 MHz \cite{ourari2023indistinguishable}. 
The inhomogeneous linewidth of the spin transitions in \ercawo{} arises from approximately 1\% variations in the g-tensor, which is larger than typical values in other Er systems~\cite{Chen2020}, and we attribute it to the strain sensitivity of the $S_4$ site~\cite{ourari2023indistinguishable}.

\begin{figure}[H]
    \makebox[\textwidth][c]{\includegraphics[]{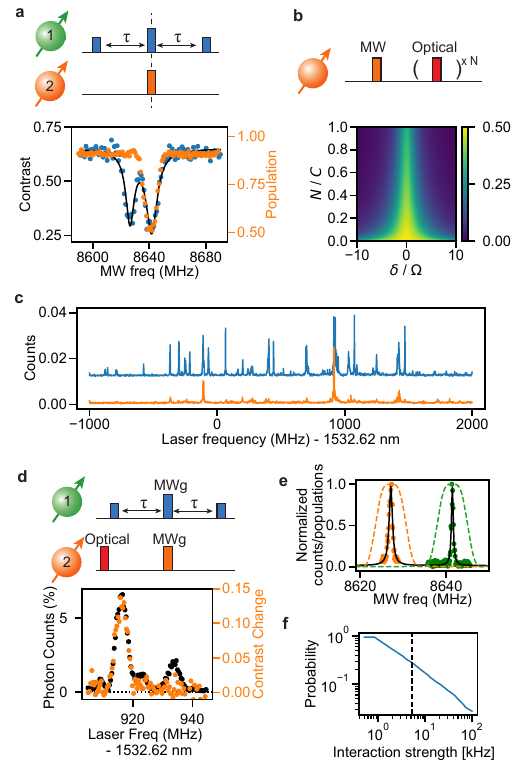}}%
    \caption{Search for interacting \er{} pairs 
    \textbf{a,} DEER spectroscopy (blue markers) reveals two dips in spin coherence, fitted by the solid line. One of the dips overlap with the Er-1 optically detected magnetic resonance (ODMR) (orange markers), and another dip suggests the existence of an interacting \er{} neighbor, Er-2. 
    \textbf{b,} The selectivity of MW assisted PLE spectroscopy with respect to the MW frequency (MW detuning compared to Rabi frequency, $\delta/\Omega$) can be adjusted by varying the number of optical pulses per MW remixing pulse $N$ compared to the cyclicity $C$ of the optical transition. For non-selective experiments ($N/C \ll 1$), fluorescence is still observed when the MW detuning from the spin transition $\delta$ is larger than the drive Rabi frequency, $\Omega$.
    \textbf{c,} Non-selective MW assisted PLE with $N=1$ reveals most of the ions in the cavity (blue trace), while a selective experiment with $N=50$ only excites one ion with the corresponding MW frequency (orange trace). The two peaks for the selective ($N=50$) case are the A and B optical transitions of the same ion, Er-2. 
    \textbf{d,} Exciting Er-2 before the DEER sequence shelves the spin and changes the DEER contrast (orange). The change in contrast overlap with the PLE of Er-2 (black). 
    \textbf{e,} ODMR on Er-1 (green markers) and on Er-2 (orange markers), fitted by Lorentzians (black solid line) yield linewidths of $246$~kHz and $488$~kHz of Er-1 and Er-2 respectively. The calculated population change under a $\pi$-pulse (dotted trace) shows negligible crosstalk with probability of accidental spin flip less than 0.1\%.
    \textbf{f,} Cumulated distribution function of the largest Ising interacting strength of an ion with its neighbors (blue) yields a probability of $26\%$ to have an interacting neighbor with $J \geq 5.40$~kHz Ising interaction (black dashed line).
}
    \label{fig:SI_IonProperties}
\end{figure}

A large number of ions are observed within the cavity, but the average ion separation of 140~nm (planar implantation with $5\times10^9\,\text{cm}^{-2}$) means that most ions are not interacting coherently.
To identify individual ions with strong coupling to neighbors, we use double electron-electron resonance (DEER) spectroscopy, consisting of the Hahn sequence on Er-1 and a resonant MW pulse on the potential interacting \er{} (Extended Fig.~\ref{fig:SI_IonProperties}a, top). 
The DEER spectrum of Er-1 (Extended Fig.~\ref{fig:SI_IonProperties}a, bottom) shows two peaks: one corresponds to the Er-1 ground state spin transition frequency, and the second suggests the existence of a nearby interacting ion, Er-2. 
To find the optical transition of Er-2, we perform MW-assisted PLE, where microwaves at the MWg frequency are used to remix dark states during PLE spectroscopy. 
For a sufficiently high microwave power or pulse repetition rate, this remixing is effective for all ions in the sample, in spite of the inhomogeneity in the g-tensor. 
However, reducing the repetition rate of MW pulses makes the remixing more selective, and allows for recording a PLE spectrum from only ions with MWg near the MW drive frequency (Extended Fig.~\ref{fig:SI_IonProperties}b). 
By performing a selective MW assisted PLE experiment at the MW frequency obtained from the DEER experiment, we find the optical transition frequencies of an ion with the desired MW transition frequency (Fig.~\ref{fig:SI_IonProperties}c).
To confirm that this optical transition indeed belongs to the interacting ion, we excite this optical transition before the DEER experiment to shelve the population in the excited state, and find that the DEER signal is reduced (Extended Fig.~\ref{fig:SI_IonProperties}d). 
This confirms the assignment of the optical line to the interacting ion observed in the DEER experiment.

The separation of the optical transitions (200 MHz) and spin transitions (14 MHz) between Er-1 and Er-2 allows for individual control.
To avoid crosstalk of the MW pulses while improving the robustness against spin dephasing, Hermite functions are used as the pulse envelope \cite{Warren1984} with negligible crosstalk (less than 0.1\% to accidentally flip the other spin for each $\pi$-pulse) (Extended Fig.~\ref{fig:SI_IonProperties}e). 
However, this limits the usable Rabi frequency to around 10~MHz. 
For future works, further pulse engineering, \emph{i.e.} gradient ascent pulse engineering or optical AC stark shift, can be used to implement faster gate while maintaining small crosstalk \cite{khaneja2005optimal,Chen2020}.

From the measured interaction strength ($J=5.40(3)$~kHz), we can estimate the distance between the ions. 
However, this single-angle measurement cannot pinpoint the ion–ion distance, and it only sets bounds. 
From the maximum dipolar interaction possible, we obtain an upper-bound separation of 42~nm. 
A lower bound of 10 nm is deduced from the absence of flip-flop processes ($T_{1}=2.5(2)$ s), given the 14 MHz spin transition frequency mismatch between the ions. 
Given the planar implantation density of $5\times10^9\,\text{cm}^{-2}$ into a layer with thickness around 20 nm, the probability for any ion to have a neighbor with an interaction strength comparable to the one in this paper is $26\%$ (Extended Fig.~\ref{fig:SI_IonProperties}f).

\subsection{Multi-ion initialization and readout}\label{SI:erReadout}
\begin{figure}[H]
    \makebox[\textwidth][c]{\includegraphics[]{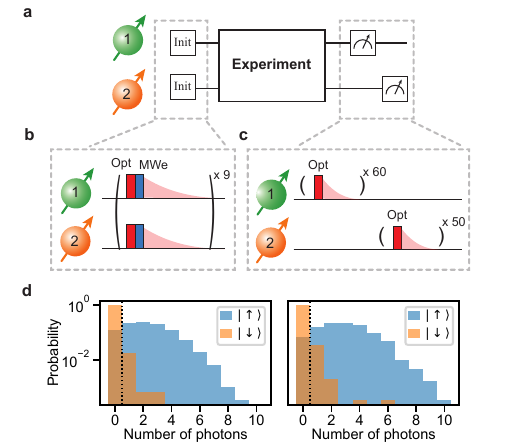}}%
    \caption{Performing experiments on two ions 
    \textbf{a,} Circuits for running experiments on two ions consist of initializations, experiments, and readouts. 
    \textbf{b,} Each initialization round consists of optical excitations followed by MW pulses to pump the population out of the cycling transition. 
    \textbf{c,} Readout consists of repeated optical excitations on the cycling transitions to collect enough photons for thresholding. The excitations on the two ions are performed sequentially for temporal separation of their photons.
    \textbf{d,} The histograms of photon detection events for reading out two ions, thresholding with one photon detection event gives readout fidelities of 0.95 and 0.94 on Er-1 (left) and Er-2 (right) respectively.
}
    \label{fig:SI_IonReadouts}
\end{figure}

The difference in the optical and spin transition frequencies of Er-1 and Er-2 allows us to control them independently (Extended Fig.~\ref{fig:SI_IonReadouts}a). 
The initializations of the two ions are performed simultaneously, where two tones of optical and MWe pulses are used to excite both ions and then pump the populations out of the cycling transitions (Extended Fig.~\ref{fig:SI_IonReadouts}b).
The readouts of the two ions are performed sequentially, so that the collected photons are separated temporally (Extended Fig.~\ref{fig:SI_IonReadouts}c). 
By optically exciting Er-1 60 times to collect enough photon detection events for thresholding, we achieve a readout fidelity of $F_{\text{readout}}^{(1)}=0.95$ (Extended Fig.~\ref{fig:SI_IonReadouts}d, left) \cite{Chen2020,ourari2023indistinguishable}. 
Similarly, a readout fidelity of $F_{\text{readout}}^{(2)}=0.94$ is achieved for Er-2 with 50 optical excitations (Extended Fig.~\ref{fig:SI_IonReadouts}d, right). 
To perform readout correction, we first use this single-shot readout to obtain the expectation value, \emph{i.e.} $\langle Z_i\rangle$, and then scale the result by $(2\times F_{\text{readout}}^{(i)}-1)^{-1}$. 
For the correlation results, \emph{i.e.} $\langle Z_1Z_2\rangle$, we scale the expectation by $(2 F_{\text{readout}}^{(1)}-1)^{-1}(2 F_{\text{readout}}^{(2)}-1)^{-1}$.

\subsection{\er{}-\er{} gate construction}\label{SI:er_gate}

\begin{figure}[H]
    \makebox[\textwidth][c]{\includegraphics[]{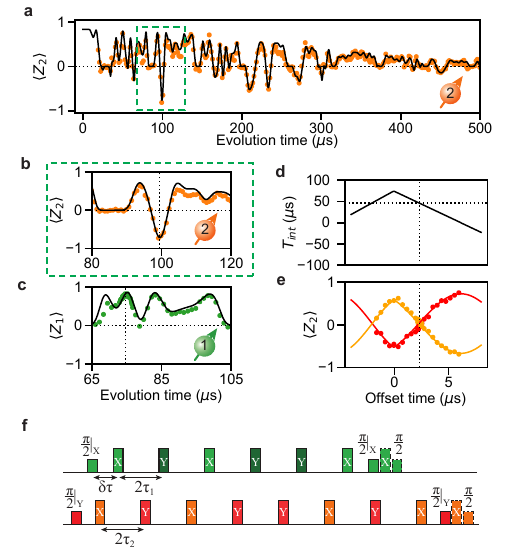}}%
    \caption{\er{}-\er{} entangling operations 
    \textbf{a,} Er-2 spin coherence under the XY-8 sequence (markers) shows ESEEM features due to surrounding nuclear spins. Correlated Cluster Expansion (CCE) simulation of two measured strongly coupled nuclear spins (Extended Tab.~\ref{Tab:SI_NucProperties}) and a randomly generated weakly coupled bath (black solid line) show agreement with the data \cite{ourari2023indistinguishable}.
    \textbf{b,} Zoomed-in Er-2 spin coherence under the XY-8 sequence (markers) shows a contrast of $p^{(2)}_{(XY-8)}=-0.86$ at the evolution time of $16\tau=99.3$~$\mu$s (vertical dashed line) that is used for the entangling gate.
    \textbf{c,} Er-1 spin coherence under the XY-6 sequence (markers) shows a contrast of $p^{(1)}_{(XY-6)}=0.87$ at the evolution time of $12\tau=74.6$~$\mu$s (vertical dashed line).
    \textbf{d,} Shifting the time offset $\delta\tau$ between two overlapping dynamical decoupling sequences changes the effective interaction time $T_{\text{int}}$. The horizontal dashed line corresponds to the required interaction time for maximal entanglement.
    \textbf{e,} Er-2 spin coherence when sweeping $\delta\tau$ after being initialized into two different bases (red and orange markers) matches the calculated result based on $T_{\text{int}}$ (red and orange lines). The maximal entanglement is reached when $\delta\tau=2.3$~$\mu$s, where no contrast is observed when only measuring Er-2. 
    \textbf{f,} The pulse sequence for CZ gate is placed between $\pi/2$-pulses at the beginning and end of the sequence to create and measure the Bell state. The final $\pi/2$-pulses in the dashed box are the analysis pulses for reading out different bases, and extra $\pi$-pulses are used for refocusing.
}
    \label{fig:SI_ErGates}
\end{figure}

The magnetic-dipolar interaction between the ions is assumed to be purely Ising-type because of the large spin frequency difference (14~MHz) and the long spin $T_{1}=2.5(2)$ s, precluding interaction terms that flip the spin.
This Ising interaction naturally allows for the construction of the control-Z (CZ) gate. 
To accumulate a two-qubit phase of $\pi$ required for maximal entanglement, the interaction time must be $(4J)^{-1}=46.3$~$\mu$s. 
This required interaction time is significantly longer than the electron spin free precession time $T_{2}^{*}\approx1\,\mu$s. 
The coherence time can be extended further using dynamical decoupling, but care must be taken to choose the inter-pulse spacing to avoid ESEEM features, which result from unwanted entanglement between the electron spins and the surrounding nuclei.
For example, the contrast of XY-8 on Er-2 is shown in Extended Fig.~\ref{fig:SI_ErGates}a. 
The coherence time is extended to 360~$\mu$s, with modulations at particular inter-pulse spacings that are resonant with the precession frequencies of the nearby nuclear spins. 
To avoid unwanted entanglement, it is necessary for the contrast to be +1 (no phase accumulation on the electron from the nuclear interaction) or -1 ($\pm\pi$ phase accumulation on the electron from the nuclear spin interaction). 
For Er-2, we choose the latter operating point with a total evolution time of $99.3$~$\mu$s. 
For Er-1, we use an XY-6 and operate with the total evolution time of $74.6$~$\mu$s where the contrast is +1. 
The effective interaction time $T_{\text{int}}$ can be controlled by adjusting the offset time between the start of each pulse sequence (Fig.~\ref{fig:Fig2}a, Extended Fig.~\ref{fig:SI_ErGates}d,e). 
While there are many possible parameters for DD sequences that yield a correct gate, it is practically helpful to ensure that the $\pi$-pulses on the ions do not temporally overlap, and that the effective interaction time $T_{\text{int}}$ is a large fraction of the total spin evolution time.
We also note that operating Er-2 at a time where the contrast is -1 results in a byproduct single-qubit rotation on the nuclear spin of $\pm\pi$ (or a relative phase of $2\pi$).

The CZ gate is placed between $\pi/2$-pulses to construct the Bell state (Extended Fig.~\ref{fig:SI_ErGates}f). To measure the Bell state in $Z_1Z_2$ basis, we directly perform single-shot readout on each ion right after the pulse sequences. To measure the state in $X_1X_2$ or $Y_1Y_2$ bases, we need to apply extra $\pi/2$-pulses to change bases, during which the Bell state $(\ket{\uparrow\uparrow}+\ket{\downarrow\downarrow})/\sqrt{2}$ may suffer from dephasing. Therefore, another set of $\pi$ pulses are inserted for refocusing.

\subsection{Er-Er gate fidelity estimation}\label{SI:gate_fidelity}
Electron spin decoherence is the major source of error for all gates in this work. 
To estimate the \er{}-\er{} gate fidelity limited by spin coherence, we directly measure the XY-6 and XY-8 contrast on two ions with evolution times $74.6$~$\mu$s and $99.3$~$\mu$s as used in the Er-Er CZ gate (Extended Fig.~\ref{fig:SI_ErGates}b,c). 
The absolute values of contrasts are $p^{(1)}_{(XY-6)}=0.87$ and $|p^{(2)}_{(XY-8)}|=0.86$ for Er-1 and Er-2 respectively. 
Note that this measurement includes both spin decoherence and pulse errors (limited by the electron free precession time $T_{2}^{*}\approx1$ $\mu$s compared to the $\pi$-pulse duration of $50$ ns) in the sequence.
Then, the final contrast in all different basis in the Bell measurement can be estimated to be the product of two contrasts $\langle X_1X_2\rangle=\langle Y_1Y_2\rangle=\langle Z_1Z_2\rangle=p^{(1)}_{(XY-6)}|p^{(2)}_{(XY-8)}|=0.748$. 
This corresponds to the fidelity of $F=(1+\langle X_1X_2\rangle+\langle Y_1Y_2\rangle+\langle Z_1Z_2\rangle)/3=0.81$, slightly larger than the measured fidelity of $F=0.76$. 
The disagreement in fidelity might result from changes in pulse calibrations between the XY-N sequences and gate implementations. 

The same CZ gate in the remote readout experiment, on the other hand, agrees with the electron spin coherence. 
Because only Er-1 was in a superposition state in this experiment, we only consider its dephasing effect, its pulse errors, and its single-shot readout fidelity. For Er-2, we only consider its pulse errors. 
This results in $F=(1+p^{(1)}_{(XY-6)}F_{\text{readout}}^{(1)}(1-p^{(2)}_{\text{error}}/2))/2=0.89$, where $F_{\text{readout}}^{(1)}=0.94$ is the single-shot readout fidelity of Er-1 (Extended Fig.~\ref{fig:SI_IonReadouts}d) and $p^{(2)}_{\text{error}}=0.08$ is the probability of pulse errors on Er-2, extracted from Fig.~\ref{fig:Fig2}d. This is in agreement with the measured remote readout fidelity of 0.88.
 
\subsection{Characterizing the \w{} nuclear spin}\label{SI:nuc_properties}
\begin{figure}[H]
    \makebox[\textwidth][c]{\includegraphics[]{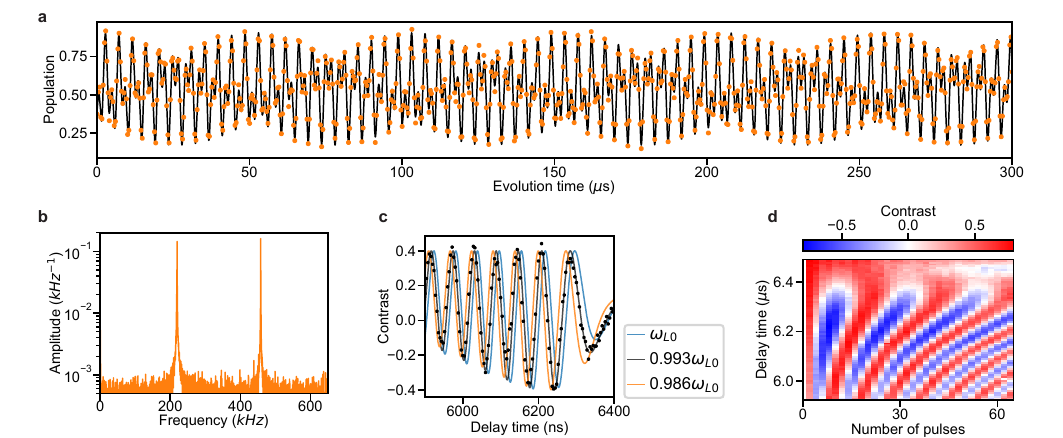}}%
    \caption{Characterizing the nuclear spin
    \textbf{a,} Nuclear spin free precession (markers) shows the sum of two oscillations at frequencies $\omega_+$ and $\omega_-$ fitted by the solid line.
    \textbf{b,} The Fourier transform of the free precession shows two peaks corresponding to $\omega_+=458$~kHz and $\omega_-=219$~kHz.
    \textbf{c,} The measured ESEEM features in XY-96 on Er-2 (markers) show disagreement with simulation with $\omega_{L0}$, the ideal Larmor frequency, from which we estimated the actual Larmor frequency to be around $0.993\omega_{L0}$.
    \textbf{d,} Varying the number of pulses and the inter-pulse spacing of the pulse sequence on the electron spin results in a chevron pattern on the nuclear spin evolution. The chevron pattern is distorted because the parallel hyperfine coupling strength is comparable to the Larmor frequency, $A_{\parallel}\approx\omega_{L}$.
}
    \label{fig:SI_NucProperties}
\end{figure}

The \w{} nuclear spin Hamiltonian (Eq.~\ref{eq:Nuc_H}) consists of the nuclear spin Larmor frequency $\omega_{L}$, and the parallel and perpendicular hyperfine coupling strengths $A_{\parallel}$ and $A_{\perp}$. 
The Larmor frequency can be roughly estimated from the \w{} nuclear spin g-factor, $g_{n}=0.2356$, in the external magnetic field $B_{\text{ext}}=790$ G, and the hyperfine parameters can be measured using the nuclear spin Ramsey experiment as in Fig.~\ref{fig:Fig4}a and in Extended Fig.~\ref{fig:SI_NucProperties}a,b \cite{uysal2023coherent}: 
Varying the time between the two CZ gates shows the sum of two nuclear spin free precessions $\omega_{\pm}$ (Extended Fig.~\ref{fig:SI_NucProperties}a). 
From the Fourier transform of the Ramsey in Extended Fig.~\ref{fig:SI_NucProperties}b, the two precession frequencies are measured to be $\omega_{+}=458$~kHz and $\omega_{-}=219$~kHz. 
Then, the hyperfine parameters can be calculated to be $A_{\parallel}=287$~kHz and $A_{\perp}=163$~kHz.

The ESEEM features from the nuclear spins, especially for higher order dynamical decoupling sequences, show small disagreements from the simulation (Extended Fig.~\ref{fig:SI_NucProperties}c). 
The disagreements can be matched by assigning slightly different Larmor frequencies to different nuclear spins, and a similar behavior was also reported in \cite{Wang2023}. 
Therefore, we design gates based on a phenomenological model where the Larmor frequencies can vary for each nuclear spin. 
To estimate the Larmor frequency precisely, we fit both the precession frequencies (Extended Fig.~\ref{fig:SI_NucProperties}b) as well as the ESEEM features in the XY-96 sequence (Extended Fig.~\ref{fig:SI_NucProperties}c), and extract the Larmor frequency to be $0.7\%$ lower than expected.

Using these procedures, we characterize one nuclear spin near Er-1 and two nuclear spins near Er-2, which are listed in the Extended Tab.~\ref{Tab:SI_NucProperties}. 
Nuc-1 near Er-2 is used as the ancilla memory in the main text.

\subsection{Nuclear spin coherence}\label{SI:nuc_coh}
\begin{figure}[H]
    \makebox[\textwidth][c]{\includegraphics[]{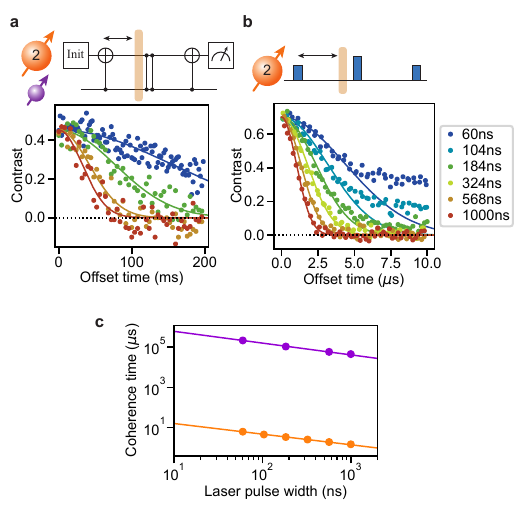}}%
    \caption{
    Optically induced dephasing
    \textbf{a,} \w{} nuclear spin Hahn experiment with an additional optical pulse drawn as the shaded orange rectangle. The optical pulse is off-resonant with all the \er{} ions but is still in the photonic cavity linewidth. The nuclear spin coherence decays when the timing of the optical pulse approaches the refocusing operations.
    \textbf{b,} Er-2 \er{} electron spin Hahn experiment with additional off-resonance optical pulse. The \er{} coherence decays when the optical pulse is approaching the refocusing pulse, as also reported in \cite{uysal2025}.
    \textbf{c,} Coherence time of Er-2 (orange markers) and the \w{} spin (purple markers) as a function of laser pulse width. The two fitted curves (orange and purple lines) are offset by a factor of $3.2\times10^4$, roughly matching the ratio of $|A|/\omega_{\text{MWg}}=2.7\times10^4$.
}
    \label{fig:SI_OptDep}
\end{figure}

The nuclear spin free precession time in the Ramsey experiment is limited by the g-tensor fluctuations of the nearby Er-2. 
As studied in the previous work, optical illumination induces local electric field fluctuations, which can lift the degeneracy of the \er{} g-tensor in $aa$-plane for electric field along the crystal $c$-axis \cite{mims1965electric,uysal2025}. 
In Extended Fig.~\ref{fig:SI_OptDep}a,b, we observe optically induced dephasing for both Er-2 and the nuclear spin. 
As the optical pulse is brought closer to the refocusing pulse of the Hahn sequence, we observe a drop in coherence in both cases, which can be explained by the optically induced dephasing mechanism, modulating the g-tensor of the \er{} ion.
For the \er{} ion, the frequency fluctuation can be expressed as $\Delta f_{e}=\omega_{\text{MWg}}\Delta g/g$, where $\omega_{\text{MWg}}$ is the Er-2 ground state spin transition frequency and $\Delta g/g$ is the ratio of change in the effective g-factor. 
For the \w{} nuclear spin, we expect the frequency fluctuation to rather go as $\Delta f_n= |A|\Delta g/g$, where $|A|$ is the magnitude of the hyperfine vector. 
In Extended Fig.~\ref{fig:SI_OptDep}c, we compare the measured dephasing times and find that they roughly differ by a factor of $|A|/\omega_{\text{MWg}}$, in agreement with expected frequency fluctuations.

\subsection{Constructing \er{}-Nuclear spin CZ gates}\label{SI:nuc_gate}
To construct the entangling gates between Er-2 and the nuclear spin, we use the knowledge of the full Hamiltonian to calculate the nuclear spin evolution under the dynamical decoupling sequences on the electron. 
Any XY-N sequences with even number of $\pi$-pulses can be decomposed into concatenated XY-2 sequences (the $\pi$-pulse angles on electron does not change the nuclear spin evolution), under which the nuclear spin evolves as:
\begin{equation}
    V_{\pm}=U_{\pm}(\tau)U_{\mp}(2\tau)U_{\pm}(\tau)=e^{-i\alpha\mathbf{q}_{\pm}/2},
\end{equation}
where $U_{\pm}(t)=e^{-i\omega_{\pm}t\mathbf{m}_{\pm}/2}$ is the nuclear spin evolution in each inter-pulse spacing. 
The net effect of the three rotations is a conditional rotation along the axes $\mathbf{q}_{\pm}$, depending on the initial electron spin state, by the same angle $\alpha$. 
To form a CZ gate, we require that $\mathbf{q}_{\pm}$ be anti-parallel, $\mathbf{q}_{+}=-\mathbf{q}_{-}$, so that the nuclear spin rotates by exactly $\pm\alpha$ along the same axis depending on the electron spin state. 
This requirement determines the resonance condition for the inter-pulse spacing $\tau$,
\begin{equation}
    \cot\left(\frac{\omega_{+}\tau}{2}\right)\cot\left(\frac{\omega_{-}\tau}{2}\right)=\cos\left(\frac{A_{\perp}\omega_{L}}{\omega_{+}\omega_{-}}\right).
\end{equation}
The solution $\tau_0$ always exists, and the corresponding rotation angle $\alpha_0$ can be calculated accordingly. 
In the case of the nuclear spin in the main text, the inter-pulse spacing is $\tau_0=6.208$~$\mu$s, and the rotation is $\alpha_0\approx\pi/4$ along the axes $\pm\mathbf{z}^\prime$ (Fig.~\ref{fig:Fig3}bc). 
Varying the number of pulses and the inter-pulse spacing results in a chevron pattern on the nuclear spin evolution when the Larmor frequency is much larger than the parallel hyperfine coupling strength ($\omega_{L}\gg A_{\parallel}$) \cite{uysal2023coherent}, but this nuclear spin is strongly interacting with electron ($\omega_{L}\le A_{\parallel}$), resulting in a distorted chevron pattern (Extended Fig.~\ref{fig:SI_NucProperties}d).

\subsection{GRASS method}\label{SI:GRASS}
As discussed in the main text, we introduce a method to construct arbitrary nuclear spin gates by solving an optimization problem, which we call the GRadient Ascent Sequence Search (GRASS) method. 
A large search space is available for decoupling sequences if the inter-pulse spacings are allowed to be arbitrary. 
Our method is inspired by the GRAPE method, which allows for pulse shape optimization for specific tasks \cite{khaneja2005optimal}. 
In this section we briefly discuss the cost function for this optimization problem and provide analytical expressions for its derivative.

In order to obtain a target unitary, we can define the following cost function:
\begin{equation}
    C_{T\pm}(\boldsymbol{\tau})=\left|\mathrm{Tr}\left[V_{\pm}(\boldsymbol{\tau})U^{\dagger}_{T\pm}\right]\right|^2,
\end{equation}
where $U_{T\pm}$ is a target unitary evolution for the nuclear spin conditional on the initial electron spin state, and $V_{\pm}(\boldsymbol{\tau})$ is the effective rotation of the nuclear spin under the sequence, which is a function of the inter-pulse spacing vector $\boldsymbol{\tau}=(\tau_1,\tau_2\ldots,\tau_{N})$ for a sequence of $N-1$ pulses. 
This cost-function, $C_{T\pm}$, expresses an overlap between the unitaries $V_{\pm}$ and $U_{T\pm}$, so that $C_{T\pm}(\boldsymbol{\tau})=d$, where $d$ is the dimension of the Hilbert space, will lead to $V_{\pm}=U_{T\pm}$ up-to a phase factor.

In order to perform the optimization efficiently, we calculate an analytic expression for the gradient of the cost function:
\begin{equation}
    \frac{\partial C_{T\pm}}{\partial \tau_j}=2\mathrm{Re}\left[\mathrm{Tr}\left[\frac{\partial V_{\pm}(\boldsymbol{\tau})}{\partial \tau_j}U^{\dagger}_{T\pm}\right] \mathrm{Tr}\left[V_{\pm}(\boldsymbol{\tau})U^{\dagger}_{T\pm}\right]^* \right].
\end{equation}
For convenience, we now calculate the derivative of $V_+$, noting that $V_-$ can be easily obtained by flipping $+$ and $-$ signs in the expressions. 
Recalling that the effective rotation operator are expressed as $V_{+}(\boldsymbol{\tau})=U_{(-1)^N}(\tau_N)\ldots U_{-}(\tau_2)U_{+}(\tau_1)$ and employing a first order expansion for the rotation operator, we can calculate the gradient:
\begin{equation}
    \frac{\partial V_{+}(\boldsymbol{\tau})}{\partial \tau_j} =
    U_{(-1)^N}(\tau_N)\ldots U_{(-1)^j}(\tau_j)\left(-i\omega_{(-1)^{j}}\mathbf{m}_{(-1)^j}\cdot\boldsymbol{\sigma}\right)U_{(-1)^{j-1}}(\tau_{j-1})\ldots U_{+}(\tau_1)
\end{equation}
This equation tells us that in order to compute the gradient with respect to $\tau_j$ for all $j$, we only need to evaluate each free-precession unitary $U_{\pm}(\tau_j)$ once. 
Intuitively, the derivative can be grouped into a backward propagator $U_{(-1)^N}(\tau_N)\ldots U_{(-1)^j}(\tau_j)$, a forward propagator $U_{(-1)^{j-1}}(\tau_{j-1})\ldots U_{+}(\tau_1)$, and the precession axis operator $\mathbf{m}_{(-1)^j}\cdot\boldsymbol{\sigma}$.

Similarly, we can define the cost functions to other nearby nuclear spins to avoid unwanted entanglement. 
When there are multiple nuclear spins in the solid-state environment, a coherent operation on the desired nuclear spin requires decoupling of the electron spin from the remainder of the nuclear spins. 
To decouple from the remainder nuclear spins, the following cost function can be maximized:
\begin{equation}
    C_{D}(\boldsymbol{\tau}) = \left|\mathrm{Tr}\left[V_{+}(\boldsymbol{\tau})V_{-}^{\dagger}(\boldsymbol{\tau})\right]\right|^2.
\end{equation}
The cost function $C_{D}(\boldsymbol{\tau})$ evaluates the overlap between the effective conditional rotations $V_{+}$ and $V_{-}$. 
Maximizing the overlap implies that the nuclear spin undergoes the same evolution regardless of the electron spin state, thus minimizing its entanglement. 
This nuclear spin, therefore, only experiences a known single-qubit rotation, meaning that the information inside can still be retrieved if it is also used as a memory. 
The derivative of this cost function can be calculated similarly:
\begin{equation}
    \frac{\partial C_{D\pm}(\boldsymbol{\tau})}{\partial \tau_j}=2\mathrm{Re}\left[\mathrm{Tr}\left[\frac{\partial V_{+}(\boldsymbol{\tau})}{\partial \tau_j}V_{-}^{\dagger}(\boldsymbol{\tau})+V_{+}(\boldsymbol{\tau})\frac{\partial V_{-}^{\dagger}(\boldsymbol{\tau})}{\partial \tau_j}\right] \mathrm{Tr}\left[V_{+}(\boldsymbol{\tau})V_{-}^{\dagger}(\boldsymbol{\tau})\right]^* \right],
\end{equation}
yielding an expression again in terms of $V_{\pm}(\boldsymbol{\tau})$ and its derivatives.

Another important addition to the cost function is to ensure that the pulse sequence is still a decoupling sequence for the electron spin. 
This can be achieved by ensuring that the electron spends equal time in even and odd windows of the sequence so that any frequency fluctuation that is static over the sequence is decoupled. 
The corresponding cost function can be calculated as:
\begin{equation}
    C_S(\boldsymbol{\tau}) = -\left(\sum_{k=0}^N(-1)^k\tau_k\right)^2
\end{equation}
Maximizing this cost function ensures decoupling of the static noise by requiring $\sum_{\textrm{even } k}\tau_k = \sum_{\textrm{odd } k}\tau_k$. 
The derivative can also be calculated as:
\begin{equation}
    \frac{\partial C_{S\pm}(\boldsymbol{\tau})}{\partial \tau_j}=
    (-1)^j\left(2\sum_{k=0}^N(-1)^k\tau_k\right).
\end{equation}

After the construction of all the necessary cost functions, a simple gradient ascent algorithm can then be followed in the following manner:
\begin{itemize}
    \item Calculate the gradient $\frac{\partial C(\boldsymbol\tau)}{\partial \boldsymbol\tau}$
    \item Update the inter-pulse spacing vector: $\boldsymbol{\tau}=\boldsymbol{\tau}+\epsilon\frac{\partial C(\boldsymbol\tau)}{\partial \boldsymbol\tau}$
    \item Repeat until convergence or a set number of iterations.
\end{itemize}

In our search for a sequence, we have optimized for a controlled-Y (CY) operation using the following cost function:
\begin{equation}
    C = C_{T+}^{(1)} + C_{T-}^{(1)} + C_D^{(2)} + C_S,
\end{equation}
where the target unitary operations were $U_{T+} = X$ and $U_{T-} = Z$ for the cost-functions $C_{T\pm}^{(1)}$ targeting the first nuclear spin. 
This optimization target results in an effective CY rotation together with a single-qubit gate on the nuclear spin, $\ket{\uparrow}\bra{\uparrow}\otimes X+\ket{\downarrow}\bra{\downarrow}\otimes Z=U_S\left(\ket{\uparrow}\bra{\uparrow}\otimes V+\ket{\downarrow}\bra{\downarrow}\otimes V^\dagger\right)$, where $U_S=-iI\otimes H$ is a Hadamard gate on the nuclear spin and $V=e^{iY\pi/4}$ is the CY rotation.
$C_D^{(2)}$ ensures decoupling from the second nuclear spin and the term $C_S$ ensured the decoupling condition for the electron spin from its environment. 
The hyperfine parameters used in the cost functions for the first and second nuclear spins are noted in Table~\ref{Tab:SI_NucProperties}.
In addition to the second nuclear spin, the \er{} spin is also weakly coupled to a \w{} nuclear spin bath, whose hyperfine parameters we do not measure. While the above cost function does not ensure decoupling from the bath, we can achieve decoupling by generating and testing multiple (about 200) pulse sequences obtained by performing the optimization with different initial conditions.
We finally choose the shortest sequence, with 8-pulses and nine intervals given by $\{\tau_i\}=\{1.928,\,5.392,\,7.056,\,7.874,\,7.380,\,2.926,\,4.822,\,7.116,\,2.086\}$ ($\mu$s) as also shown in Fig.~\ref{fig:Fig3}d.

\subsection{Construction of Er-Nuclear spin circuits}\label{SI:nuc_circuit}
The circuits on two \er{} ions and the nuclear spin are performed by concatenating pulse sequences on two ions (Extended Fig.~\ref{fig:SI_NucGates}). 
Because the nuclear spin precesses around two different axes $\mathbf{m}_{\pm}$ depending on the Er-2 spin state, there is no well-defined rotating frame and the nuclear spin free precession always needs to be considered during the circuits. Therefore, the sequences of the gates in a circuit are concatenated back-to-back without extra waiting times (Extended Fig.~\ref{fig:SI_NucGates}a).
Two $\pi/2$-pulses may overlap when sequences are concatenated, in which case the two pulses might be canceled or replaced by an equivalent pulse.

The full diagrams of the circuits to generate \er{}-nuclear spin Bell state are shown in Extended Fig.~\ref{fig:SI_NucGates}bcde.
The circuit to generate Bell state and measure the $Z_{1}Z_{2}$ correlation contains four SWAP operations, each consists of two CX operations because only the $Z$ basis information needs to be transferred between qubits.
The circuit to measure the $X_{1}X_{2}$ correlation, on the other hand, has one more CX operation in the last SWAP operation, because we need to swap the superposition state to Er-2 and then perform a $\pi/2$ rotation before readout.

\subsection{Er-Nuclear spin circuits fidelities}\label{SI:nuc_gate_fidelity}
The finite electron spin coherence limits the fidelity of both electron-electron and electron-nuclear spin gates. 
The errors during the gates can be estimated by performing the gate twice to remove the entanglement and then recording the remainder electron coherence. 
The CZ gate contains the XY-4 sequence with an evolution time of $49.7$ $\mu$s, during which the electron coherence is estimated to be $p_{\text{CZ}}=\sqrt{p_{\text{CZ}^2}}=0.93$. 
Measurement on two CX gates gives a single operation contrast of $p_{\text{CX}}=\sqrt{p_{\text{CX}^2}}=0.94$. 
Then, the SWAP gate fidelity and thus the retrieved state fidelity (Fig.~\ref{fig:Fig4}ab) can be estimated to be $F=(1+p_{\text{CX}}^2p_{\text{CZ}}^4)/2=0.830$, showing agreement with the measured result of $F=0.833(3)$.

\begin{figure}[H]
    \makebox[\textwidth][c]{\includegraphics[]{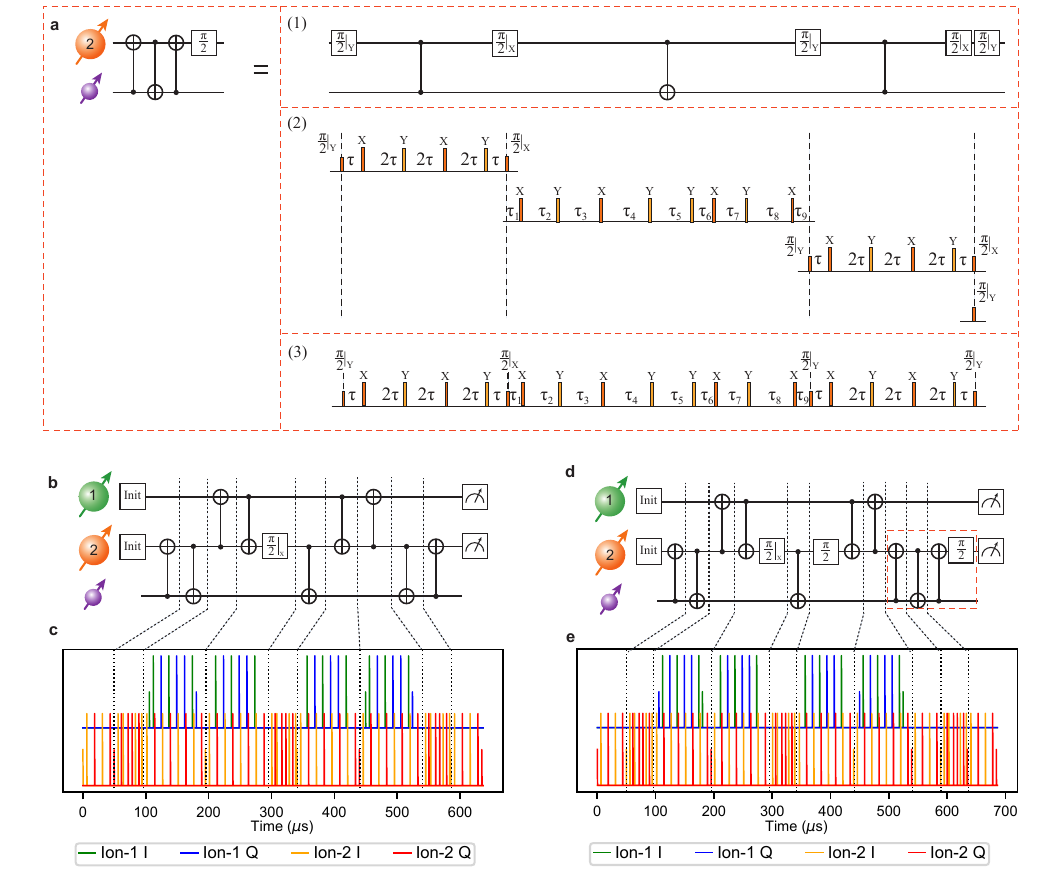}}%
    \caption{Pulse sequence for \er{}-Nuclear spin Bell state preparation and measurement 
    \textbf{a,} Schematic diagram of pulse sequence concatenation in part of the circuit in \textbf{d}. (1) The three CX operations are formed by $\pi/2$-pulses, the CZ gate, and the CX gate. (2) The CZ and CX gates are constructed from pulse sequences. (3) The pulse sequences are concatenated. The last $\pi/2$-pulse along X is skipped, because it does not affect the readout on $\ket{\pm}$ (projected to $\ket{0/1}$ by $\pi/2$ along Y).
    \textbf{b,} Circuit diagram for Bell state creation and measurement in the $ZZ$ basis.
    \textbf{c,} Pulse sequences in I/Q channels of the $ZZ$ basis measurement on Er-1 and Er-2 are shown. 
    \textbf{d,} Circuit diagram of the $XX$ basis measurement is shown. 
    \textbf{e,} Pulse sequences on Er-1 and Er-2 are shown.
    }
    \label{fig:SI_NucGates}
\end{figure}

The circuit to generate the \er{}-nuclear spin Bell state consists of CZ and CX operations between Er-2 and the nuclear spin, as well as the entangling operations between Er-1 and Er-2. 
The breakdown of the full circuit in both XX and ZZ basis is shown in Extended Fig.~\ref{fig:SI_NucGates}, which can be decomposed into electron-electron CZ, as well as electron-nuclear spin CZ and CX gates.
By only considering the periods where either of the electrons is in superposition state, the final expectation can be estimated as $\langle Z_1Z_2\rangle=p_{\text{CZ}}^2p_{\text{CX}}(p^{(1)}_{(XY-6)})^{2}|p^{(2)}_{(XY-8)}|^2=0.455$ and $\langle X_1X_2\rangle=\langle Y_1Y_2\rangle=p_{\text{CZ}}^3p_{\text{CX}}^2(p^{(1)}_{(XY-6)})^2|p^{(2)}_{(XY-8)}|^2=0.398$. 
Then, the estimated fidelity is $F=(1+\langle X_1X_2\rangle+\langle Y_1Y_2\rangle+\langle Z_1Z_2\rangle)/4=0.56$, slightly larger than the measured fidelity of $F=0.52(2)$.

\newpage

\newpage

\newpage

\newpage

\newpage

\newpage

\newpage

\begin{table}[H]
    \centering
     \begin{tabular}{|c|c|c|c|c|} 
         \hline
         & Coupled ion & $A_{\parallel}$ (kHz) & $A_{\perp}$ (kHz) & $\omega_{L}$ (kHz) \\ \hline
         Nuc-1 &  Er-2 & 287 & 163 & 142 \\ \hline
         Nuc-2 & Er-2 & 168 & 130 & 144 \\ \hline
         Nuc-3 & Er-1 & 54 & 134 & 144 \\ \hline
     \end{tabular}
    \caption{Hyperfine parameters for the nuclear spins near Er-1 and Er-2. Nuc-1 is used in the text as the ancilla memory.}
    \label{Tab:SI_NucProperties}
\end{table}
 
\bibliographystyle{unsrt}
\bibliography{library}

\end{document}